\documentclass[aps,prl,twocolumn,preprintnumbers,superscriptaddress,dblfloatfix,nofootinbib]{revtex4-1}
\usepackage[utf8]{inputenc}
\usepackage{lmodern}
\usepackage{amsmath,amssymb}
\usepackage{graphicx}
\usepackage{url}
\usepackage{color}
\usepackage{enumitem}
\usepackage{subcaption}
\usepackage[dvipsnames]{xcolor}
\usepackage[colorlinks=true,breaklinks=true]{hyperref}
\hypersetup{allcolors=[rgb]{0.0 0.0 0.6},linkcolor=[rgb]{0.75 0.05 0.05}}
\usepackage{bm}
\usepackage{epsfig}
\usepackage{amsmath}
\usepackage{amssymb}
\usepackage{slashed}
\usepackage{color}
\usepackage{accents}
\usepackage[dvipsnames]{xcolor}
\usepackage[colorlinks=true,breaklinks=true]{hyperref}
\hypersetup{allcolors=[rgb]{0.0 0.0 0.6},linkcolor=[rgb]{0.75 0.05 0.05}}

\newcommand{\exclude}[1]{}

\begin{document}

\preprint{IPMU24-0022}

\title{The possibility of multi-TeV secondary gamma rays from GRB221009A}

\author{Oleg Kalashev}
\affiliation{Laboratory of Astrophysics, Ecole Polytechnique Federale de Lausanne, CH-1015, Lausanne, Switzerland}
\affiliation{Institute for Nuclear Research of the Russian Academy of Sciences,
60th October Anniversary Prospect 7a, Moscow 117312, Russia}

\author{Felix Aharonian}
\affiliation{
 Dublin Institute for Advanced Studies, School of Cosmic Physics, 31 Fitzwilliam Place, Dublin 2, Ireland}
 \affiliation{
Max-Planck-Institut f¨ur Kernphysik, Saupfercheckweg 1, 69117 Heidelberg, Germany}
\affiliation{Yerevan State University, 1 Alek Manukyan St., Yerevan 0025, Armenia}
\author{Warren Essey} 
\affiliation{Department of Physics and Astronomy, University of California, Los Angeles \\ Los Angeles, California, 90095-1547, USA} 
\author{Yoshiyuki Inoue}
\affiliation{
Department of Earth and Space Science, Graduate School of Science, Osaka University, Toyonaka,
Osaka 560-0043, Japan}
\affiliation{Interdisciplinary Theoretical \& Mathematical Science Program (iTHEMS), RIKEN, 2-1 Hirosawa, Saitama 351-0198, Japan}
\affiliation{Kavli Institute for the Physics and Mathematics of the Universe (WPI), UTIAS \\The University of Tokyo, Kashiwa, Chiba 277-8583, Japan}

\author{Alexander Kusenko} 
\affiliation{Department of Physics and Astronomy, University of California, Los Angeles \\ Los Angeles, California, 90095-1547, USA}
\affiliation{Kavli Institute for the Physics and Mathematics of the Universe (WPI), UTIAS \\The University of Tokyo, Kashiwa, Chiba 277-8583, Japan}
	
\date{\today}
	
\begin{abstract}
The brightest gamma ray burst (GRB) ever observed, GRB221009A, produced a surprisingly large flux of gamma rays with multi-TeV energies, which are expected to be absorbed in interactions with extragalactic background light (EBL).    If the highest energy gamma rays were produced at the source, their spectral shape would have to exhibit a nonphysical spike even for the lowest levels of EBL.  
We show that, for widely accepted models of EBL, the data can be explained by secondary gamma rays produced in cosmic ray interactions along the line of sight, as long as the extragalactic magnetic fields along the line of sight are $10^{-16}$G or smaller, assuming 1 Mpc correlation length.   Our interpretation supports the widely held expectation that GRB jets can accelerate cosmic rays to energies as high as 10 EeV and above, and it has  implications for understanding the magnitudes of EGMFs. 

\end{abstract}
\maketitle
	

The brightest ever observed gamma ray burst (GRB) GRB~221009A~\cite{Williams:2023sfk,LHAASO:2023lkv} defies  conventional expectations and offers an opportunity to test models of production and propagation  of gamma rays, cosmic rays, and neutrinos with multimessenger observations. In this paper we show that the {\it secondary} gamma rays, previously identified as a viable explanation of the observed  hard spectra of distant blazars, can explain the most striking  features of the GRB~221009A observations.  We also explore implications for extragalactic magnetic fields (EGMF) and extragalactic background light (EBL). 

GRB~221009A represents a new class of EBL-obscured sources similar to very distant blazars which challenge the lower bounds on extragalactic background light estimations~\cite{HESS:2005ohe, Fermi-LAT:2018lqt}.  Although some exotic, yet undiscovered particles~\cite{Galanti:2022chk,Galanti:2022xok,Lin:2022ocj} or Lorentz invariance violation~\cite{Li:2022vgq,Li:2022wxc,Nakagawa:2022wwm,Finke:2022swf} could explain the high-energy spectrum of GRB~221009A, we will consider an alternative explanation based on cosmic rays accelerated in the GRB~221009A jet generating secondary gamma rays along the line of sight~\cite{Das:2022gon}, similar to the explanation of TeV signals from distant blazars~\cite{Essey:2009zg,Essey:2009ju,Essey:2010er,Essey:2010nd,Essey:2011wv,Murase:2011cy,Razzaque:2011jc,Prosekin:2012ne,Aharonian:2012fu,Zheng:2013lza,Kalashev:2013vba,Takami:2013gfa,Inoue:2013vpa,Essey:2013kma,DeFranco:2017wdr}. We will show that this hypothesis is indeed consistent with the data.  

\begin{figure*}[!ht]
\includegraphics[width=0.49\linewidth]{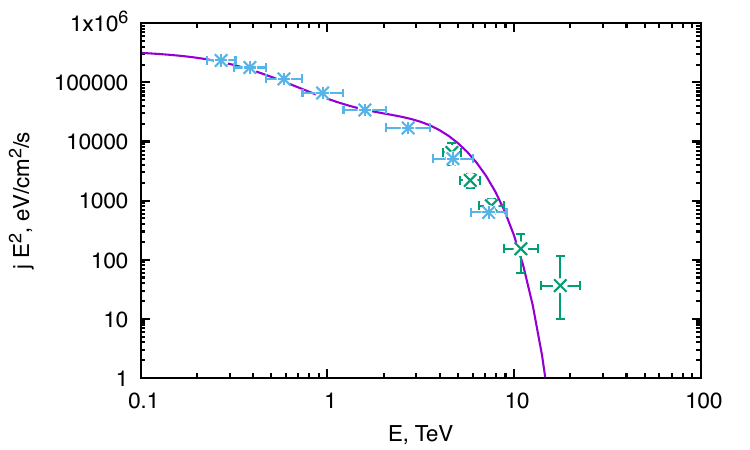}
\includegraphics[width=0.49\linewidth]
{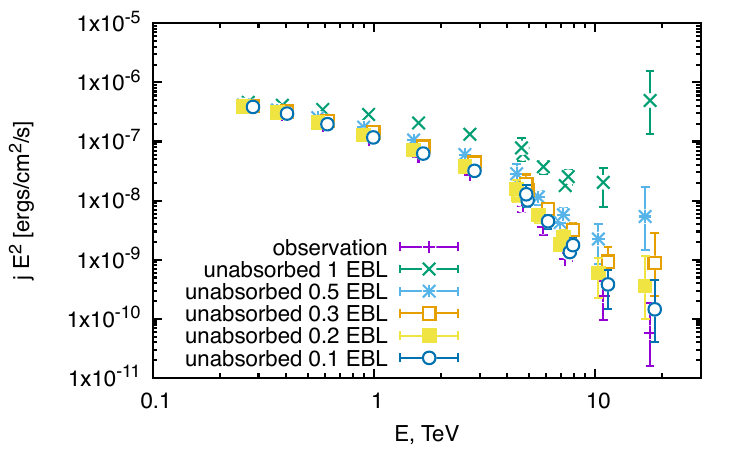}
\caption{Gamma rays produced at the source would have to overcome a growing optical depth at higher energies.  
The left panel shows the LHAASO data~\cite{LHAASO:2023lkv} and the predicted flux $j$ of gamma rays (solid line) produced at the source with a power-law spectrum extending to 10~PeV. Even with this extremely high energy cutoff, a power-law flux fails to explain the highest-energy LHAASO data point. The panel on the right shows  the reconstructed spectrum at the source, assuming the lower EBL estimate from 
Ref.~\cite{Driver:2016krv},
as well as the levels of EBL artificially scaled down by factors 0.1 to 0.5.   The spike in the gamma-ray spectrum above 10 TeV is inconsistent with models of gamma ray production.  Even if the EBL level was as low as 10-50\%, below the lower bounds based on the galaxy counts, the last data point would still required a non-physical hardening of the spectrum at the source.  }
\label{fig:primary_spectra}
\end{figure*}

\begin{figure*}[!ht]
\includegraphics[width=0.9 \linewidth]{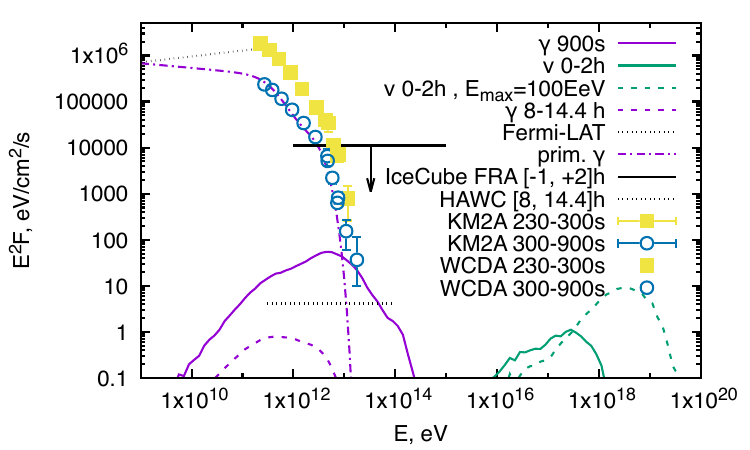}
\caption{Primary gamma rays, originating at the source with a power-law spectrum are shown by a dash-dotted line.  The secondary gamma rays and accompanying neutrinos (all flavor flux), produced in cosmic-ray interactions along the line of sight are shown by a solid lines peaked at a few TeV and a few hundred PeV, respectively.  The which explains the last bins of   the LHAASO data~\cite{LHAASO:2023lkv}. Late-time predictions for the gamma rays and neutrinos are shown by the dashed lines. Intergalactic magnetic fields were assumed to be $10^{-17}$~G, with a correlation length of 1~Mpc. Also 90\% CL upper bound on average differential neutrino all-flavor flux FRA T0 [-1:+2]h calculated from Fig. 1 of~\cite{IceCube:2023rhf} is shown}
\label{fig:spectra}
\end{figure*}
The most striking feature of GRB~221009A is the observation of gamma rays with energies above 10 TeV from an object at redshift as high as 0.151 \cite{LHAASO:2023lkv}. These data challenge conventional models of gamma-ray production at the source because the gamma rays would have to overcome attenuation due to interactions with EBL \cite{Gould:1966pza, PhysRevLett.16.479}, which would require the  spectrum at the source to have a spike at energies above TeV, even for the lowest levels of EBL consistent with galaxy counts.  In Fig.~\ref{fig:primary_spectra}, we illustrate the difficulties with explaining the data by gamma rays originating at the source.  If the gamma-ray spectrum is a power law, attenuation due to EBL results in a solid line shown in the left-hand panel of Fig.~\ref{fig:primary_spectra}.  The EBL level used in this calculation is inferred from the galaxy counts~\cite{Driver:2016krv}, which serve as a lower bound on EBL.  The last energy bin is clearly in contradiction with the prediction of the power-law model.  For this calculation we used the maximal gamma-ray energy at the source as high as 10 PeV, well above the observed data points or theoretical expectations from the acceleration models.  The short mean free path to interactions with EBL causes absorption of the highest-energy photons very close to the source. (In what follows we will consider protons, which can carry the energy away from the source and produce gamma rays much closer to the observer, evading the attenuation due to EBL.) 
If the spectrum at the source is not assumed to be a power law, one can ask what spectral shape is required to reproduce the observed data.  Such reconstructed spectrum, again using the EBL of Ref.~\cite{Driver:2016krv}, is shown as green crosses in the right-hand side panel of Fig.~\ref{fig:primary_spectra}.  
It shows a spike above 10 TeV, which could be considered a spectral
feature inconsistent with standard gamma-ray production models (see, however, Refs.~\cite{2002A&A...384..834A,2001ICRC...27I.250A}). 
One can further ask whether reducing the levels of EBL below those based on the galaxy counts (thought to be the lower limits) could allow for a monotonically decreasing spectrum at the source.  We show in the same figure the reconstructed spectra for EBL levels assumed to have the same spectral shape as in Ref.~\cite{Driver:2016krv} but multiplied by an {\it ad hoc} factor smaller than one, namely, 0.5, 0.3, 0.2, 0.1.  The spike above 10~TeV persists down to the levels of EBL that one would not be able to reconsile with galaxy counts. 

Thus the last energy bin in the LHAASO observations suggests that an explanation that is different from gamma rays produced at the source propagating through photon backgrounds might be
necessary.  As mentioned, some exotic models have been put forth, such as Lorentz invariance violation~\cite{Li:2022vgq,Li:2022wxc,Nakagawa:2022wwm,Finke:2022swf} or axion-like particles~\cite{Galanti:2022chk,Galanti:2022xok,Lin:2022ocj}.  
We will consider a different hypothesis, which was successful at explaining the highest energy gamma rays from EBL obscured distant blazars~\cite{Essey:2009zg,Essey:2009ju,Essey:2010er,Essey:2010nd,Essey:2011wv,Murase:2011cy,Razzaque:2011jc,Prosekin:2012ne,Aharonian:2012fu,Zheng:2013lza,Kalashev:2013vba,Takami:2013gfa,Inoue:2013vpa,Essey:2013kma,DeFranco:2017wdr}.

The apparent transparency of the universe to gamma rays is not unique to GRB~221009A. 
All highest energy gamma rays produced by distant sources are  expected to show a suppression in the spectrum due to pair production interactions of high-energy gamma rays with EBL \cite{Franceschini:2008tp, Finke:2009xi, Dominguez:2010bv, Gilmore:2011ks, Dwek:2012nb, Inoue:2012bk,Biteau:2015xpa,Stecker:2016fsg,Desai:2019pfl,Finke:2022uvv}. 
However, a number of blazars have spectra that extend to energies high enough~\cite{Fermi-LAT:2018lqt, MAGIC:2019ozu,VERITAS:2009lpb,MAGIC:2014vpp}
to challenge and provide potential tests of EBL models~\cite{Biteau:2015xpa,Desai:2019pfl,Greaux:2024wyc}.
It should be noted, however, that some studies of
large samples of blazar spectra find agreement between transparency predictions from EBL models and observations~\cite{Inoue:2012bk,MAGIC:2014vpp}.
Lorentz invariance violation or hypothetical new particles have been invoked to explain the blazar  data~\cite{DeAngelis:2007dqd,Simet:2007sa,Horns:2012kw,Meyer:2013fia}.
However, the blazar data can be understood if one includes the secondary gamma rays produced in line-of-site interactions of cosmic rays accelerated in the same sources~\cite{Essey:2009zg,Essey:2009ju,Essey:2010er,Essey:2010nd,Essey:2011wv,Murase:2011cy,Razzaque:2011jc,Prosekin:2012ne,Aharonian:2012fu,Zheng:2013lza,Kalashev:2013vba,Takami:2013gfa,Inoue:2013vpa,Essey:2013kma,DeFranco:2017wdr}.  This explanation works as long as the extragalactic magnetic fields are as low as $\sim 10^{-14}$G~\cite{Essey:2010nd}.  The observations  allow magnetic fields of this magnitude, while there are lower bounds $B> 2\times 10^{-14}$~G~\cite{Kandus:2010nw,Finke:2015ona,HESS:2023zwb} which depend on assumptions about the blazar duty cycles.  Of course, there can be exceptional directions along which the EGMFs are different from average, and the exceptional GRB 221009A can be seen through a window of lower EGMFs. Confirmation of secondary gamma rays hypothesis can provide information about EBL as well as the strengths of extragalactic magnetic fields along specific directions~\cite{Essey:2010nd}.  

While we do not explore all possible alternative explanations based on the models of EBL, our explanation is appealing from the point of view of the Occam’s razor. Indeed, it is widely accepted that GRB jets are likely sources of high-energy protons, and with some assumptions about the (unknown)  intergalactic magnetic fields, we can explain the data, regardless of the level or EBL.  By Occam’s razor, our explanation is quite appealing, especially considering that much more exotic models, involving axion-like particles or Lorentz invariance violation, have been proposed~\cite{Nakagawa:2022wwm,Li:2022vgq}. 

Let us summarize the lessons learned from distant blazars~\cite{Essey:2009zg,Essey:2009ju,Essey:2010er,Essey:2010nd,Essey:2011wv,Murase:2011cy,Razzaque:2011jc,Prosekin:2012ne,Aharonian:2012fu,Zheng:2013lza,Kalashev:2013vba,Takami:2013gfa,Inoue:2013vpa,Essey:2013kma,DeFranco:2017wdr} focusing on simple scaling laws for gamma rays produced at the source and along the line of sight.  The flux of primary gamma rays, produced at the source, scales with distance $d$ as~\cite{Essey:2009zg,Essey:2010er}:  
\begin{equation}
F_{\rm prim, \gamma}(d) \propto \frac{1}{d^2} \exp\{-d/\lambda_\gamma\} , 
\label{primary}
\end{equation}
where $\lambda_\gamma$ is the attenuation length due to the interactions with EBL.  The same jets that produce gamma rays in blazars (and GRBs) can accelerate protons. \footnote{In hadronic models, the accelerated protons participate in production of primary gamma rays.  In leptonic models, gamma rays can be produced without the protons' involvement, but that does not preclude the jets from production of high-energy cosmic-ray protons.} 
In addition to indirect evidence of proton acceleration in blazar jets inferred from their hard gamma-ray spectra and the need for secondary gamma rays~\cite{Essey:2009ju,Essey:2009zg,Essey:2010er}, the neutrino signal detected by IceCube from TXS~0506+056~\cite{IceCube:2018cha} supports the expectation that blazar jets can accelerate protons to very high energies.

The protons traveling along the line of sight interact with EBL and cosmic microwave background (CMB) radiation: $p\gamma_b \rightarrow p  e^+  e^-,\ p\gamma_b  \rightarrow  n\pi^+,\ p\gamma_b  \rightarrow  p\pi^0$.  These reactions, strongly boosted in the direction of the proton momentum, produce secondary gamma rays from direct pair production well as $\pi^0$ decays.
In contrast with primary gamma rays, for which the background photons cause exponential attenuation, production of secondary gamma rays has a different scaling with distance.  The background photons serve as a target for $p\gamma_b$ processes, as well as the absorbing medium.  The resulting flux scales with distance $d$ as~\cite{Essey:2010er}
\begin{equation}
 F_{{\rm sec}, \gamma}(d) \propto \frac{\lambda_\gamma}{4\pi d^2}
\left [1-e^{-d/\lambda_\gamma}  \right ]
\propto 
\left \{ 
\begin{array}{ll}
1/d, & {\rm for} \ d \ll \lambda_\gamma, \\ 
1/d^2, & {\rm for} \ d\gg \lambda_\gamma .
\end{array} \right.
\label{secondary_photons}
\end{equation}
This applies to isotropic or beamed sources, as long as the effects of the beam  broadening are small, which is the case for EGMF below 10~fG, in agreement with observational data~\citep{Kandus:2010nw,Finke:2015ona,HESS:2023zwb}.  Secondary neutrinos produced in line-of-sight interactions of photons with photon backgrounds have a flux that scales as 
\begin{equation}
F_{{\rm sec}, \nu} (d) \propto \left( F_{\rm protons} \times d \right) \propto \frac{1}{d}. 
\label{secondary_neutrinos}
\end{equation}
The 1/d scaling applies as long as the EGMFs are sufficiently small to allow the protons to remain within the angular resolution of the detector. This scaling law may be the reason why the neutrino detector IceCube has not found an association of neutrino arrival directions with nearby sources: if $F\propto 1/d^p$, the flux is  dominated by nearby sources for $p>2$ (primary neutrinos), but it is dominated by very distant sources for $p<2$ (secondary neutrinos)~\cite{Essey:2009ju,Kochocki:2020iie}.  

Both blazars and GRBs satisfy the Hillas criteria for objects capable of accelerating cosmic rays to ultrahigh energies~\cite{Kotera:2011cp}.  It is therefore reasonable to expect that protons are accelerated in GRB jets and that their interactions with cosmic photon backgrounds along the line of sight can produce secondary gamma rays with energies above TeV. 

We calculated the spectrum of secondary gamma rays numerically, taking into account deflections in the magnetic fields of the cosmic-ray protons, as well as the development of an electromagnetic cascade and deflections of the electrons and positrons, using Monte Carlo simulation CRbeam~\cite{Berezinsky:2016feh, Kalashev:2022cja} which is also available online as a part of the astrophysics data analysis platforms~\cite{CRbeamGalaxyWorkflow,CRbeamMMODA}. We assume that the primary gamma rays are produced in the GRB jet with a power-law SED of the cosmic-ray protons $dN_p/dE \propto E^{-2}$. For the EBL we used the model of Ref.~\cite{Franceschini:2017iwq}. After cascading and absorption, the SED takes the shape shown by a dotted line in Fig~\ref{fig:spectra}. There is a sharp drop in the flux of primary gamma rays around 2~TeV, beyond which the primary gamma rays are filtered out by their interactions with EBL.  

The proton injection spectrum was assumed to be a power law $dN/dE \propto E^{-2}$ with maximal energy $E_{max}=10$~EeV. Secondary gamma-ray flux produced in $p\gamma $ interactions with the  universal photon background is shown by a solid magenta line. Secondary gamma rays account for all the observed flux {\it above 10~TeV}. 

\begin{figure*}[!htb]
\centering
\begin{subfigure}{0.45\textwidth}
\includegraphics[width=\textwidth]{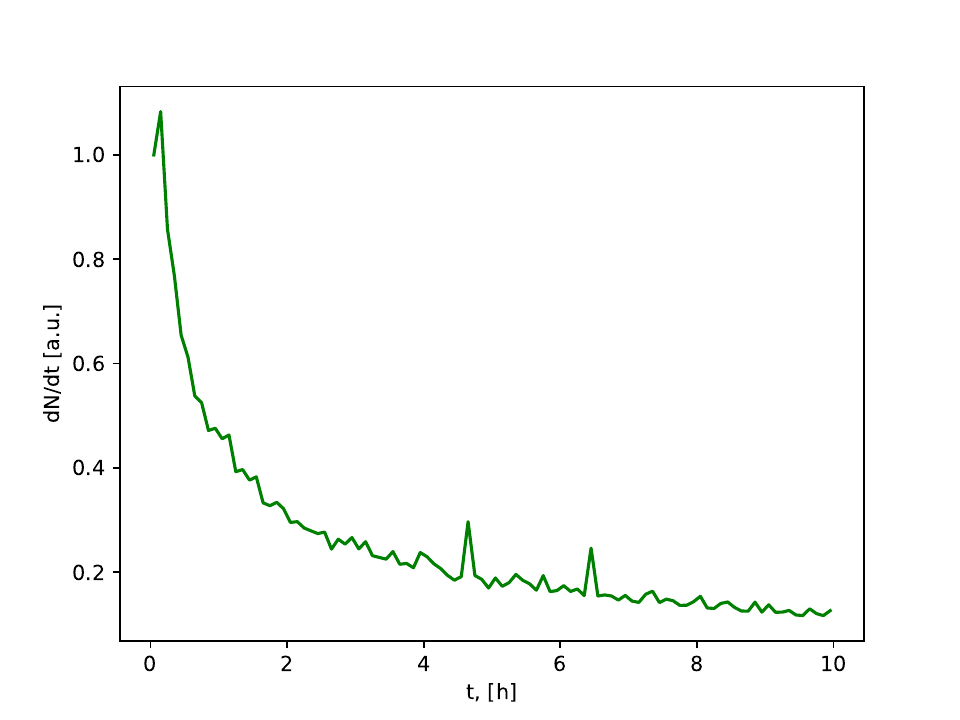}
\caption{$B=10^{-15}G$}
\end{subfigure}
\begin{subfigure}{0.45\textwidth}
\includegraphics[width=\textwidth]{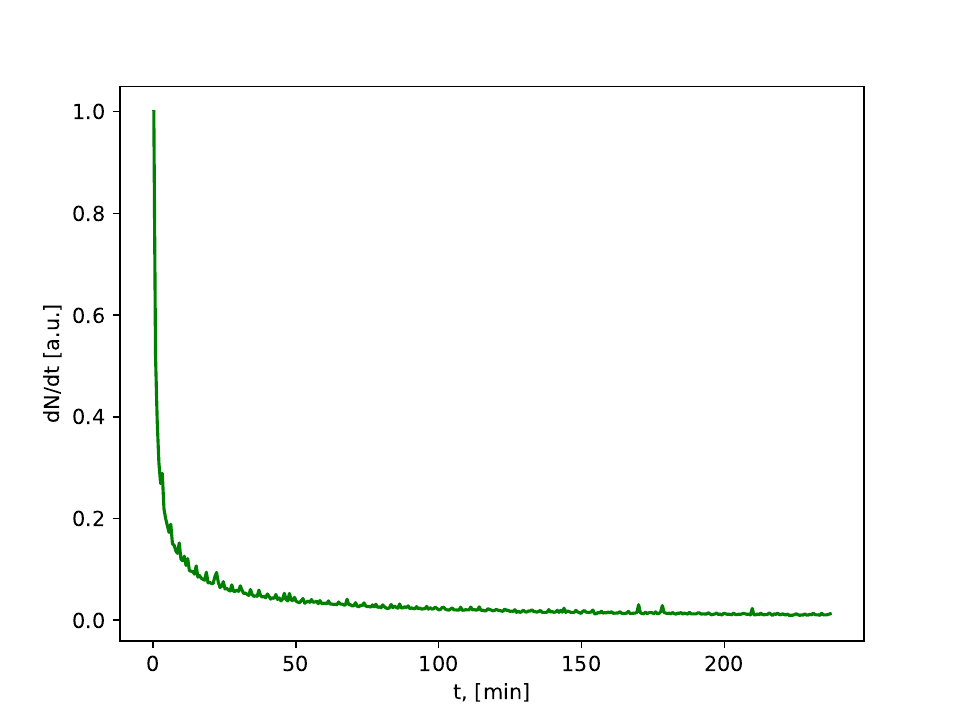}
\caption{$B=10^{-16}G$}
\end{subfigure}
\begin{subfigure}{0.45\textwidth}
\includegraphics[width=\textwidth]{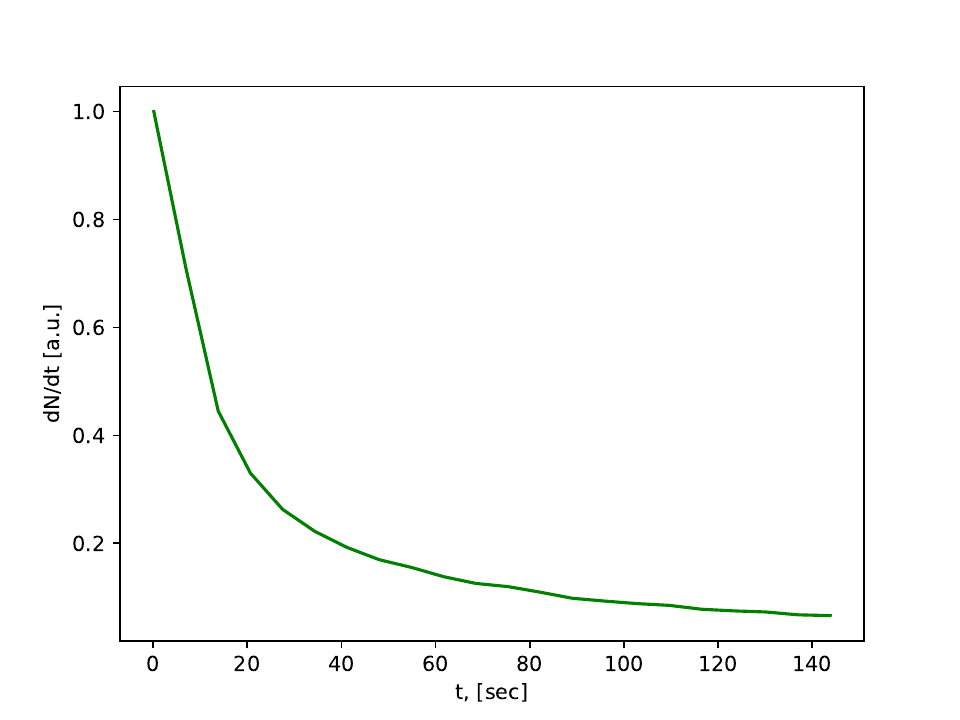}
\caption{$B=10^{-17}G$}
\end{subfigure}
\caption{Light curves of secondary $\gamma$ with energies above 10 TeV assuming source injecting protons with energy spectrum $E^{-2}$ up to 10 EeV. For the intergalactic magnetic field the correlation length of 1 Mpc was assumed and the absolute value indicated in the captions.}
\label{fig:light_curve16_17}
\end{figure*}

In addition to describing the spectrum, the secondary gamma rays must be consistent with the timeline of observations.  LHAASO observed gamma rays with energies 0.5–18 TeV~\cite{2022GCN.32677....1H} within a  time window of $\sim 2000$~s.  Magnetic fields near the source and along the line of sight cause delays in the propagation of protons as well as in the electromagnetic cascade.  
The relevant time delay, to be compared with the 2000~s observational window, is the sum of three contributions: 
\begin{equation}
 \tau_{\rm delay}=\tau_{\rm source}+\tau_{\rm p}+\tau_{\rm cascade},   
\end{equation}
where $\tau_{\rm source}$ reflects the proton delays in the local magnetic fields near the source, $\tau_{\rm p}$ accounts for the EGMF dependent delays in the proton propagation along the line of sight outside the source, and $\tau_{\rm cascade}$ is the delay in the electromagnetic cascade. 

In the case of blazars, the secondary gamma rays hypothesis implies that the magnetic fields along the line of sight must be $B \lesssim 10^{-14}$~G, assuming the correlation length of 1~Mpc~\cite{Essey:2010nd,Prosekin:2012ne}.  There is also a lower bound~\cite{Neronov:2010gir,Dermer:2010mm} on the intergalactic magnetic fields from considering sub-TeV spectra of blazars, which, taking into account possible time variability of the source, implies $B \gtrsim 10^{-18}$~G~\cite{Dermer:2010mm}.   A recent limit reported by HESS and Fermi-LAT is $B > 7\times 10^{-16}\, {\rm G}$ for duty cycle of  10~yrs.

In the host galaxy of GRB~221009A, the delays of the outgoing proton are caused by the galactic magnetic fields, which we assume to be $B_{\rm gal}\sim 10^{-6}$~G with a typical correlation length $L_{\rm gal}\sim 10$~pc. Assuming the proton with energy $E\sim 100$~EeV crosses distance $d_{\rm gal} \sim 0.1$~kpc typical of the disk thickness, it incurs a delay~\cite{Dermer:2008cy} 
\begin{equation}
\tau_{\rm source}\sim 400\, {\rm s}
\left (\frac{d}{1\, {\rm kpc}}
\frac{L_{\rm gal}}{10 \,\rm {pc}}
\right)^{3/2}
\left (\frac{B}{1 \,\rm {\mu G}}
\frac{100 \, \rm {EeV}}{E}
\right)^{2},
\end{equation}
which is smaller than the 2000~s observational time scale.  In addition to the turbulent magnetic fields, there are also regular magnetic fields in spiral galaxies, but their effect is more uncertain due to the uncertainty in the location of the GRB and the unknown configuration of the regular magnetic fields.  
The location of GRB 221009A in its host galaxy is uncertain.  HST and JWST observations~\cite{Blanchard:2023zls,Levan:2023doz} indicate that the host galaxy is an edge-on system, and the afterglow is seen at 0.25 arcsec  from the center of the image of the host galaxy.  
However, GRBs typically occur in star-forming regions, which are often found in the spiral arms of galaxies, rather than in the galactic center~\cite{Michalowski:2015bwa,deUgartePostigo:2024yat,Thone:2024myn}
If GRB 221009A occurred on the outskirts of the host galaxy, on the side of the observer, the magnetic field strength along the line of sight pointing outward from the galaxy can be  significantly lower than the micro-Gauss fields usually assumed for the regular magnetic fields inside a spiral galaxy. In contrast, the turbulent magnetic fields are produced locally by supernovae that accompany the star formation associated with the presence of large stars needed for GRBs.   
 
The remaining contributions to the arrival time delay were calculated numerically for the light curves shown in Fig.~\ref{fig:light_curve16_17} and in Table~\ref{tab:energetics}, which show consistency of the expected delays with the observational time window of $\sim 2000$~s for the EGMF amplitude of $10^{-16}$ G or smaller, assuming 1 Mpc magnetic field correlation length.  

As shown in Fig.2, the highest energy bin, which is not explained by the primary gamma rays contains the data from the time interval $300 {\rm s} < t < 900 {\rm s}$.  Therefore, we can assume that the protons leave the source after a delay of the order of $300$~s, and the relevant secondary gamma-ray flux is accumulated during the following 600~s.  
Under this assumption, the required (isotropic) source luminosity in protons $L_{\rm iso}$ with energy above 1 EeV and the fraction $f_{\rm E\gamma\, sec}$ of the proton energy in secondary gamma rays with energies above 10 TeV were calculated for models with $E_{max}=10$ EeV and $E^{-2}$ injection spectrum.  We assume that the predicted flux of secondary gamma rays fits the last energy bin of KM2A data between 300 and 900 seconds (see Fig.~\ref{fig:spectra}). 
The results are presented in Table~\ref{tab:energetics}.


\begin{table}[]
    \centering
    \begin{tabular}{|c|c|c|c|c|}
    \hline
\ \ \ $B$, G \ \ \    &  \ \ \  $ \tau_{\rm obs}/s$   \ \ \    &  \ \ \  $ \tau_{\rm med}/s$   \ \ \   &  \ \ \  $ f_{\rm E\gamma\, sec}$   \ \ \ &  $\log (L_{\rm iso}/{\rm erg})$  \\ \hline 
  $10^{-17}$  & $\lesssim 18$ &  $144$ &  0.0039   & 53.5 \\ \hline
    $10^{-16}$  & $\lesssim 72$ & $1.44\times 10^4$ &  0.00062   & 54.3 \\ \hline
      $10^{-15}$  & 5500 &  $1.44\times 10^6$ &  $3.6\times 10^{-5} $   & 55.5 \\ \hline
    \end{tabular}
\caption{\small Time interval 
$\protect \tau_{\rm obs}$ corresponding to the  flux decrease by a factor $\protect 1/e$ and a median delay $\protect \tau_{\rm med}$ for secondary gamma rays with energy above {10 TeV}, as well as the fraction $\protect f_{\rm E\gamma}$ of the initial energy that reaches an observer in the form of 10-TeV secondary gamma rays within 600 seconds, and corresponding isotropic source luminosity in cosmic rays $\protect L_{\rm iso}$ above 1 EeV required to explain the observations, depending on the intergalactic magnetic field assumed. A realistic beaming factor reduces the actual energy requirements \cite{Rudolph:2022dky} by factor 10-100.   }
    \label{tab:energetics}
\end{table}


We have shown that secondary gamma rays produced in cosmic-ray interactions along the line of sight provide a viable explanation of the highest energy data reported by LHAASO.  This appears to confirm that GRBs can accelerated protons to very high energies, as was already pointed out in Ref.~\cite{Das:2022gon}.  Furthermore, the observed gamma rays are consistent with the expected time delays, as long as the EGMFs along the line of sight  are below $10^{-16}$G, assuming 1 Mpc correlation length.  Our results are not based on a comprehensive statistical analysis, which is left for future studies.  
The small magnetic fields in the voids may be vestiges of primordial seed fields, which could open a new window on the early universe cosmology with future gamma-ray data.  Importantly, our results show that there is no conflict between the LHAASO observations and the models of EBL, because secondary gamma rays can account for the highest-energy data.

The work of A. K. was supported by the U.S. Department of Energy (DOE) Grant No. DE-SC0009937;  by World Premier International Research Center Initiative (WPI), MEXT, Japan; and by Japan Society for the Promotion of Science (JSPS) KAKENHI Grant No. JP20H05853. YI is supported by NAOJ ALMA Scientific Research Grant Number 2021-17A; by World Premier International Research Center Initiative (WPI), MEXT; and by JSPS KAKENHI Grant Number JP18H05458, JP19K14772, and JP22K18277.


\bibliography{main}


\end{document}